\documentclass[aps,prb, twocolumn,showpacs]{revtex4-1}
\usepackage{amsmath,amssymb,latexsym}
\usepackage{graphicx}
\usepackage{bm}

\usepackage[normalem]{ulem}

\usepackage{color}

\begin{document}

\title{Quantum many-body dynamics of the Einstein-de Haas effect}
\author{J.~H.~Mentink}
\email{j.mentink@science.ru.nl}
\affiliation{Radboud University, Institute for Molecules and Materials, Heyendaalseweg 135, 6525 AJ, Nijmegen, the Netherlands}
\author{M.~I.~Katsnelson}
\affiliation{Radboud University, Institute for Molecules and Materials, Heyendaalseweg 135, 6525 AJ, Nijmegen, the Netherlands}
\author{M.~Lemeshko}
\affiliation{Institute of Science and Technology Austria, Am Campus 1, 3400 Klosterneuburg, Austria}

\date{\today}

\begin{abstract}
In 1915, Einstein and de Haas and Barnett demonstrated that changing the magnetization of a magnetic material results in mechanical rotation, and vice versa. At the microscopic level, this effect governs the transfer between electron spin and orbital angular momentum, and lattice degrees of freedom, understanding which is key for molecular magnets, nano-magneto-mechanics, spintronics, and ultrafast magnetism. Until now, the timescales of electron-to-lattice angular momentum transfer remain unclear, since modeling this process on a microscopic level requires addition of an infinite amount of quantum angular momenta. We show that this problem can be solved by reformulating it in terms of the recently discovered angulon quasiparticles, which results in a rotationally invariant quantum many-body theory. In particular, we demonstrate that non-perturbative effects take place even if the electron--phonon coupling is weak and give rise to angular momentum transfer on femtosecond timescales.
\end{abstract}

\maketitle

\section{Introduction}
The concept  of  angular momentum is ubiquitous across physics, whether one deals with nuclear collisions,  chemical reactions, or formation of galaxies. In the microscopic world, quantum rotations are described by non-commuting operators. This makes the angular momentum theory extremely involved, even for systems consisting of only a few interacting particles, such as electrons filling an atomic shell or protons and neutrons composing a nucleus~\cite{VarshalovichAngMom}. In condensed matter systems, exchange of angular momentum between electrons' spins and a crystal lattice governs the Einstein-de Haas~\cite{EinsteindeHaas15de} and Barnett~\cite{BarnettPR15} effects. These effects play a key role in magnetoelasticity~\cite{VlasovJETP1964}, in the physics of molecular and atomic magnets~\cite{CaleroPRL05,GaraninPRX2011,GanzhornNatCom16,DonatiScience16}, nano-magneto-mechanical systems~\cite{WallisAPL06,KovalevPRL05,JaafarPRB2009,Tejada2010PRL,KeshtgarPRB17}, spintronics~\cite{KovalevPRB2007,MatsuoPRL11,MatsuoJPSJ17}, and ultrafast magnetism~\cite{BeaurepairePRL96, KoopmansPRL2005, KirilyukRMP10, Dornes2019}.

If approached from first principles, describing angular momentum transfer in condensed-matter systems represents a seemingly intractable problem, since it involves couplings between an essentially infinite number of angular momenta of all the electrons and nuclei in a solid. As a result, although several models of spin--lattice coupling have been developed~\cite{VanVleckPR40,Callen1963,MelcherPRL72,FeddersPRB77,Mishchenko1997, ChudnovskyPRB05, ZhangPRL14, GaraninPRB15, FahnleJSNM17}, they either solve the problem only partially (i.e.\ by ignoring the orbital dynamics of electrons) or do not account for the overall rotational invariance of the microscopic Hamiltonian. Moreover, while non-perturbative effects of electron--phonon coupling have been shown to play an important role in solid-state systems, most notably in the theory of polarons \cite{AppelPolarons} and in the microscopic theory of BCS superconductivity \cite{BardeenPR1957}, none of the existing theories of angular momentum transfer have been applied beyond the perturbative regime. As a result, over 100 years after their discovery, a fully quantum mechanical microscopic description of the Einstein-de Haas and Barnett effects remains elusive. In particular, due to these limitations existing theories cannot describe how fast angular momentum can be transferred between electronic and lattice degrees of freedom.

\begin{figure}[b]
\includegraphics[width=\columnwidth]{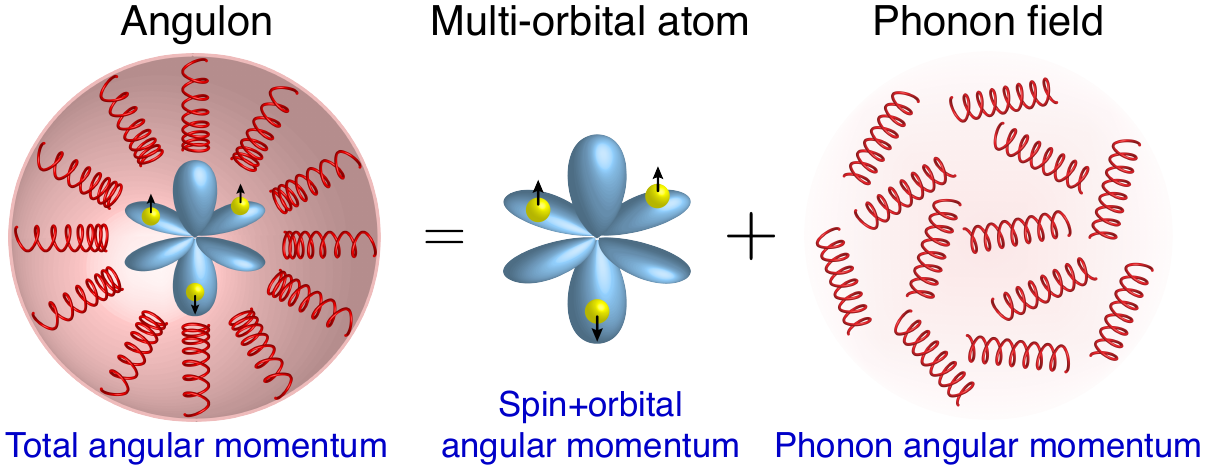}
\caption{
{\bf Angulon quasiparticle in solid-state systems.} A localized magnetic impurity exchanging angular momentum with lattice excitations can be described as the angulon quasiparticle, characterized by total (electrons+phonons) angular momentum.
\label{angulon}}
\end{figure}

Here, we introduce a conceptually novel approach to angular momentum transfer in solids, which relies on casting both electron and lattice degrees of freedom, and {-- most importantly  --} the coupling between the two, directly in the angular momentum basis. This results in a fully rotationally invariant quantum many-body theory that treats both electron spin and orbital angular momenta as well as phonon angular momentum on an equal footing. Remarkably, despite the fact that this problem involves coupling between an infinite number of angular momenta, it can be solved in closed form in terms of the angulon quasiparticle, a concept that was recently discovered in molecular physics~\cite{LemeshkoDroplets16}. In the solid-state context, the angulon represents a many-electron atom dressed by a cloud of lattice excitations carrying angular momentum, see Fig.~\ref{angulon}. This quasiparticle approach not only captures perturbative effects such as the renormalization and broadening of well-known low-frequency properties, but also makes it straightforward to take non-perturbative effects into account.

We emphasize that taking a phonon-dressed many-electron atom as building block represents a key step beyond conventional theories of electron-phonon coupling. Such theories usually account for phonons on top of an electronic Hamiltonian involving non-local interactions (electron hopping, static crystal fields), which were recently argued to dominate the ultrafast angular momentum dynamics in electron-only theories since they break rotational symmetry\cite{TowsPRL15,Dewhurst2018}. However, by construction, electron-only theories fail to describe how, and how fast angular momentum is transferred from electronic to lattice degrees of freedom. Moreover, even when accounting for such non-local interactions, rotational invariance of the system as a whole should still be conserved.

In this paper, as the first application of our formalism, we focus on the local angular momentum transfer between electrons and phonons, which is of key importance to reveal the shortest possible timescale of the Einstein de Haas effect \cite{KoopmansPRL2005,Dornes2019}, and for which the angulon building block alone is sufficient. If required, electron hopping and crystal fields can be introduced on top of such a building block. This, however, should not alter the qualitative behavior of the electron-phonon system described in this paper. Interestingly, already at this level, we predict qualitatively novel non-perturbative effects taking place even if the electron--lattice coupling is weak. These features arise at high energies and therefore enable transfer between electron spins and phonons at ultra short timescales. 

\section{The Microscopic model}

To illustrate our approach, we consider a microscopic Hamiltonian, $\hat H=\hat H_\text{e} + \hat H_\text{p} + \hat H_\text{ep}$, where $\hat H_\text{e}$ accounts for the electronic degrees of freedom, $\hat H_\text{p}$ describes the phonons, and $\hat H_\text{ep}$ captures the electron--phonon coupling. For concreteness, as $\hat H_\text{e}$ we take the multi-orbital  (in this case, three-orbital) Hubbard-Kanamori Hamiltonian describing localized paramagnetic atoms \cite{GeorgesARCMP13} with an additional spin-orbit coupling term. We explicitly consider the limit where the electronic degrees of freedom are completely localized on the atom and describe the atomic Hamiltonian as
\begin{align}
\label{eq:He}
\hat H_\text{e}= \hat H_N - 2J_\text{H}\hat{\mathbf{S}}^2 - \frac{J_\text{H}}{2}\hat{\mathbf{L}}^2 + \xi\, \hat{\mathbf{L}}\cdot\hat{\mathbf{S}}.
\end{align}
Here $\hat H_N = \hat{N}(\hat{N}-1)(U-3J_\text{H})/2 + 5J_\text{H}\hat{N}/2 $, where $\hat N$ is the total electron number operator, $U$ and $J_\text{H}$ parametrize the direct and exchange Coulomb interactions, respectively, and $\xi$ gives the the spin-orbit coupling strength. $\hat{\mathbf{L}}$ and $\hat{\mathbf{S}}$ are many-electron operators for the orbital and spin angular momentum, respectively, and we use $\hbar \equiv 1$ such that the parameters $U,J_\text{H}$, and $\xi$ have the dimension of energy. In an isolated atom, angular momentum $\hat{\mathbf{J}} =\hat{\mathbf{L}}+\hat{\mathbf{S}}$ is conserved and $\hat H_\text{e}$ is diagonal in the many-electron states, $|\Gamma\rangle=|NLSJM_J\rangle$ \cite{RudzikasBook2007,IrkhinPSSB1994}, where $M_J$ is the projection of $\mathbf{J}$ onto the laboratory-frame $z$-axis. 

For the sake of simplicity, we describe the lattice degrees of freedom by considering an isotropic elastic solid whose excitations are acoustic phonons as described by the Hamiltonian
\begin{equation}
\label{eq:Hp}
\hat H_\text{p}=\sum_{k\lambda\mu s}\omega_{ks}\,\hat{b}^\dagger_{k\lambda\mu s}\hat{b}_{k\lambda\mu s},
\end{equation}
with a linear dispersion, $\omega_{ks}=c_sk$, $c_s$ being the speed of sound, $s$ the polarization index, and $k=|\mathbf{k}|$. In Eq.~\eqref{eq:Hp} we have used the angular momentum representation for the creation and annihilation operators, $\hat{b}^\dagger_{k\lambda\mu s}$ and $\hat{b}_{k\lambda\mu s}$, where $\lambda$ and $\mu$ give the phonon angular momentum and its projection onto the $z$-axis, respectively\cite{LemSchmidtChapter}. The boson operators in the $\{ k, \lambda, \mu\}$ basis are connected to the operators in the Cartesian representation, $\hat{b}^\dagger_{\mathbf{k}s}$ and $\hat{b}_{\mathbf{k}s}$, with $\mathbf{k} \equiv \{k_x, k_y, k_z\}$, as follows:
\begin{align}
\hat{b}^\dagger_{\mathbf{k}s}=\frac{(2\pi)^{3/2}}{k}\sum_{\lambda\mu}\hat{b}_{k\lambda\mu s}^\dagger \text{i}^{\lambda}Y^*_{\lambda\mu}(\Omega_k),\label{bdagtransf}
\end{align}
In this angular momentum representation, each phonon carries angular momentum $\lambda$ with projection $\mu$ and $\hat{H}_\text{p}$ is diagonal in the basis $|k\lambda\mu s\rangle$.  The total angular momentum of phonons with a given polarization is then defined  by summing all excited phonons according to their occupations. For the three different components of the total phonon angular momentum we get the following expression: 
\begin{align}
\hat{\mathbf{\Lambda}}_s=\sum_{k\lambda\mu\mu'}\hat{b}_{k\lambda\mu s}^\dagger\boldsymbol{\sigma}_{\mu\mu'}^\lambda \hat{b}_{k\lambda\mu' s},
\end{align}
where $\boldsymbol{\sigma}^\lambda$ is the vector of matrices fulfilling the angular momentum algebra in the representation of angular momentum $\lambda=0,1,2\ldots$\cite{SchmidtLemPRX16} Hence, the total phonon angular momentum defined in this way is composed of non-spherical excitations of the elastic solid (e.g., $p,d,f$-waves for $\lambda=1,2,3$).
 
The next step is to formulate the electron--phonon coupling, $\hat{H}_\text{ep}$, in a rotationally invariant way. Here we outline the main steps of this derivation, and provide further details in Appendix~\ref{a1}. Our starting point is the general Hamiltonian describing density--density interactions between electrons and ions of the lattice, 
\begin{align}
\label{eq:Hepgen}
\hat{H}_\text{ep}=\int \!\!d\mathbf{x}\int \!\!d\mathbf{r}\,\,\hat{\Psi}^\dagger(\mathbf{x})\hat{\Phi}^\dagger(\mathbf{r})V(\mathbf{x},\mathbf{r})\hat{\Psi}(\mathbf{x})\hat{\Phi}(\mathbf{r}),
\end{align}
where $\hat{\Psi}(\mathbf{x})$ and $\hat{\Phi}(\mathbf{r})$ are field operators for electrons and nuclei, respectively. Microscopically, the two-body interaction, $V(\mathbf{x},\mathbf{r})$, stems from the Coulomb interaction between electrons and nuclei, which is obviously rotationally invariant, $V(\mathbf{x},\mathbf{r}) \equiv V(|\mathbf{x}-\mathbf{r}|)$. Hence, rotational invariance is implied and the task is to describe excitations between different angular momentum states of electrons and phonons due to such an isotropic interaction. For this purpose we first expand the interaction in spherical harmonics, $Y_{lm}(\Omega)$:
\begin{align}\label{Vscalar}
V(\mathbf{x},\mathbf{r})=\sum_{lm}V_l(x,r)Y^*_{lm}(\Omega_x)Y_{lm}(\Omega_r),
\end{align}
Second, considering electrons localized around the nuclei, we expand $\hat{\Psi}^\dagger(\mathbf{x})=\sum_{j}\hat{\psi}^\dagger_j(\mathbf{x}-\mathbf{r}_j)$ and construct the local field operators $\hat{\psi}^\dagger_j(\mathbf{x})$ from a complete set of atomic orbitals:
\begin{align}\label{atomicorbitals}
\hat{\psi}^\dagger(\mathbf{x})=\sum_{\lambda\mu,\sigma}\rho_{\nu\lambda}(x)Y^*_{\lambda\mu}(\Omega_x)\chi^\dagger_{\sigma}\, \hat{c}^\dagger_{\lambda\mu \sigma},
\end{align}
Here $\chi^\dagger_\sigma$ is a Pauli spinor and $\hat{c}^\dagger_{\lambda\mu \sigma}$ is the electron creation operator. The indices $\nu,\lambda,\mu,\sigma$ are the principal and orbital angular momentum quantum numbers and the projections of orbital and spin quantum numbers, respectively. Finally, we introduce phonons by expanding $V(\mathbf{x},\mathbf{r})$ in small displacements and subsequent transformation to the spherical phonon basis. In the resulting Hamiltonian integration over electronic and
nuclear angles can be performed analytically. Here we present the result for the case  in which (phonon-mediated) hopping between different atoms in the lattice is neglected:
\begin{align}
\label{eq:Hep}
\hat H_{\text{ep},\lambda_1}^{\text{loc}}&=\sum_{\mu_1\mu_2}\sum_{k\lambda\mu}U_\lambda(k)\frac{\text{i}}{2}\left(1+(-1)^\lambda\right)\\
&\quad\times\left[-A^{\lambda_1\mu_1}_{\lambda\mu,\lambda_2\mu_2}\hat{b}_{k\lambda\mu} +  (-1)^\mu A^{\lambda_1\mu_1}_{\lambda-\mu,\lambda_2\mu_2}\hat{b}^\dagger_{k\lambda\mu}\right]\nonumber\\
&\quad\times\!\!\!\!\sum_{NS\mathit{\Sigma}\atop LL'MM'}\!\!\!\!N\,W^{LL'S}_{MM'\mu_1\mu_2} \hat X(NLMS\mathit{\Sigma},NL'M'S\mathit{\Sigma})\nonumber,
\end{align}
It is important to note that, first, only terms with $\mathbf{k}\cdot \mathbf{e}_{s}(\mathbf{k})\neq0$ (with $\mathbf{e}_{s}(\mathbf{k})$ the polarization vector)  survive, as follows from the expansion of $V(\mathbf{x},\mathbf{r})$ to first order in nuclear displacements. This implies that only longitudinal phonons contribute in the case of an isotropic elastic solid. Second, in Eq.~\eqref{eq:Hep} we introduced the $\hat X$-operators \cite{IrkhinPSSB1994} (or Hubbard operators \cite{HubbardPRSA1965}), $\hat X(\Gamma,\Gamma')=|\Gamma\rangle\langle \Gamma'|$, that describe the transitions between many-electron states due to the terms $\hat{c}^\dagger_{\lambda_1\mu_1\sigma}\hat{c}_{\lambda_1\mu_2\sigma}$. Here  $A^{\lambda_1\mu_1}_{\lambda\mu,\lambda_2\mu_2}$ captures the selection rules for single-electron excitations due to phonons (see Appendix \ref{a1}) and $W^{LL'S}_{MM'\mu_1\mu_2}$ determines the allowed transitions between many-electron terms with different orbital angular momenta, $LM\neq L'M'$. In contrast, $S\mathit{\Sigma}=S'\mathit{\Sigma}'$, since the electron--phonon coupling does not depend on spin $S$ and its projection, $\mathit{\Sigma}$. We emphasize that $W^{LL'S}_{MM'\mu_1\mu_2}$ is based upon the exact solution of the many-electron problem, which takes into account all allowed electronic transitions with $N$ and $N\pm1$ electrons. The coupling strength is determined by $U_\lambda(k)$ that originates from the radial integrals. Explicit formulas for $U_\lambda(k)$ and $W$ are given in Appendix~\ref{a1}. Third, we stress that although we started from a spherically symmetric Coulomb interaction, the charge distribution of the atomic orbitals is not spherically symmetric. As a result, the coupling between different non-spherical electron distributions induced by phonons leads to angular momentum transfer. Indeed, when including only $s$-orbitals no transfer takes place, so one needs to have asymmetric $p$-, $d$-, or $f$-orbitals. At the same time, rotational invariance is preserved, \textit{i.e.} simultaneous rotation of both electron and phonon subsytems leaves $H_\text{ep}^\text{loc}$ unchanged.

The full Hamiltonian, $\hat H=\hat H_\text{e} + \hat H_\text{p} + \hat H^\text{loc}_\text{ep}$, is rotationally invariant and is therefore diagonal in the basis of a given total angular momentum, $|\mathsf{J}\mathsf{M}_\mathsf{J}\rangle$.

In addition, it exhibits striking similarities with the one used to describe molecules rotating in superfluids, which were recently found to form so-called angulon quasiparticles \cite{SchmidtLem15, SchmidtLemPRX16, LemeshkoDroplets16}. Instead of mechanical rotation of a molecule, here we deal with orbital angular momentum of electrons. Lattice phonons, on the other hand, play the role of superfluid excitations. The anisotropic molecule--helium interaction, in turn, is replaced with the rotationally invariant electron--phonon coupling, Eq.~\eqref{eq:Hep}, derived here from microscopic principles. Inspired by this analogy, in what follows we make use of the angulon concept in order to understand  angular momentum transfer in solid-state systems.

The key advantage of casting the problem in terms of angulons is that it allows for a drastic simplification. The latter, in turn, enables studying non-perturbative effects  based on a transparent variational ansatz. By analogy with the molecular angulon, we construct an ansatz featuring all possible single-phonon excitations allowed by angular momentum conservation in the subspace of a given number of electrons, $N$:
\begin{align}
&|\psi_{\mathsf{JM_J}}\rangle = Z^{1/2}_{\mathsf{JM_J}}|LSJM_J\rangle|0\rangle \label{variationalpsi}\\
&\qquad + \sum_{k\lambda\mu \atop lm}\beta^{\mathsf{JM_J}}_{k\lambda l}\sum_{M\mathit{\Sigma}} C^{JM_J}_{LM,S\mathit{\Sigma}}C^{LM}_{lm,\lambda\mu} \hat{b}^\dagger_{k\lambda\mu}|0\rangle|lmS\mathit{\Sigma}\rangle\nonumber,
 \end{align}
A similar ansatz has been previously shown to provide a good approximation to the energies of polarons \cite{CombescotPRL08} and angulons \cite{BighinPRL18}, even far beyond the weak-coupling regime considered in this paper. In Eq.~\eqref{variationalpsi}, $Z^{1/2}_{\mathsf{JM_J}}$ and $\beta^{\mathsf{JM_J}}_{k\lambda l}$ are variational parameters which are determined by minimizing $\langle \psi_{\mathsf{JM_J}}| H- E|\psi_{\mathsf{JM_J}}\rangle$. This yields the equation, $E = E_{\mathsf{JM_J}} - \Sigma_{\mathsf{JM_J}}(E)$, from which the variational ground-state energy, $E$, is determined self-consistently. Here $E_{\mathsf{JM_J}}$ is the energy of the many-electron state without phonons and $\Sigma_{\mathsf{JM_J}}(E)$ plays the role of a self-energy describing the effect of electron--phonon interactions:
\begin{align}
\Sigma_{\mathsf{JM_J}} (E)&=\sum_{k\lambda l} \frac{U_\lambda(k)^2 Q_{\lambda l}^2}{E^\mathsf{JM_J}_{\lambda l}- E +\omega_k},
\end{align}
where $Q_{\lambda l}$ are matrix elements that determine the allowed transitions to electronically excited states, $E^\mathsf{JM_J}_{\lambda l}$, due to phonons with angular momentum $\lambda$, which are given in an explicit form in the Appendix~\ref{a2}. Non-perturbative effects described below originate from the energy, $E$, in the denominator of $\Sigma_\mathsf{JM_J}(E)$. These effects do not take place in conventional second-order perturbation theory, which is recovered by replacing $\Sigma_\mathsf{JM_J}(E)\rightarrow \Sigma_\mathsf{JM_J}(E_{\mathsf{JM_J}})$.

Moreover, the quasiparticle approach enables the study of angular momentum transfer in response to a time-dependent magnetic field, as described by the Zeeman term:
\begin{align}
\hat H_Z(t)=\mu_\text{B}\mathbf{B}(t)\cdot\left(g_L\hat{\mathbf{L}}+g_S\hat{\mathbf{S}}\right),
\end{align}
In this case, we search for a solution based on the time-dependent variational principle \cite{McLachlanMolPhys1964,JackiwPhysLettA1979}. Following Ref.\cite{KatsnelsonJPhysC1984}, we write $|\Psi(t)\rangle=e^{-\text{i}Et}\sum_{\mathsf{M}_\mathsf{J}}|\psi_{\mathsf{JM_J}}(t)\rangle$.   Next, for each  $\mathsf{M}_\mathsf{J}$,  we use the variational ansatz \eqref{variationalpsi} with time-dependent parameters, $Z^{1/2}_{\mathsf{JM_J}}(t)$ and $\beta^{\mathsf{JM_J}}_{k\lambda l}(t)$, which are determined by minimizing $\langle \Psi(t)| \text{i}\partial_t - H- H_Z(t)|\Psi(t)\rangle$.  {Crucially,} this variational approach also gives rise to non-perturbative effects in the dynamical response. That is, in addition to the perturbative effects that give rise to phonon dressing of states with different $\mathsf{M_J}$, qualitatively new features appear.

Within the quasiparticle picture this can be understood as follows. Due to the static phonon dressing, an external field can trigger virtual transitions to atomic states with $J'\neq J$, where $J$ is the ground-state angular momentum of the isolated atom. Without phonon coupling such excitations are obviously forbidden due to selection rules. Moreover, in the presence of electron--phonon coupling, these electronically excited states can decay by emitting phonons. This 
can either lead to (i) emergence of quasi-bound states of the quasiparticle itself, where reduced angular momentum of the electrons is balanced by increased phonon angular momentum or (ii) give rise to incoherent scattering of phonons. Both effects are captured by our theory.

\begin{figure}[t]
\includegraphics[width=\columnwidth]{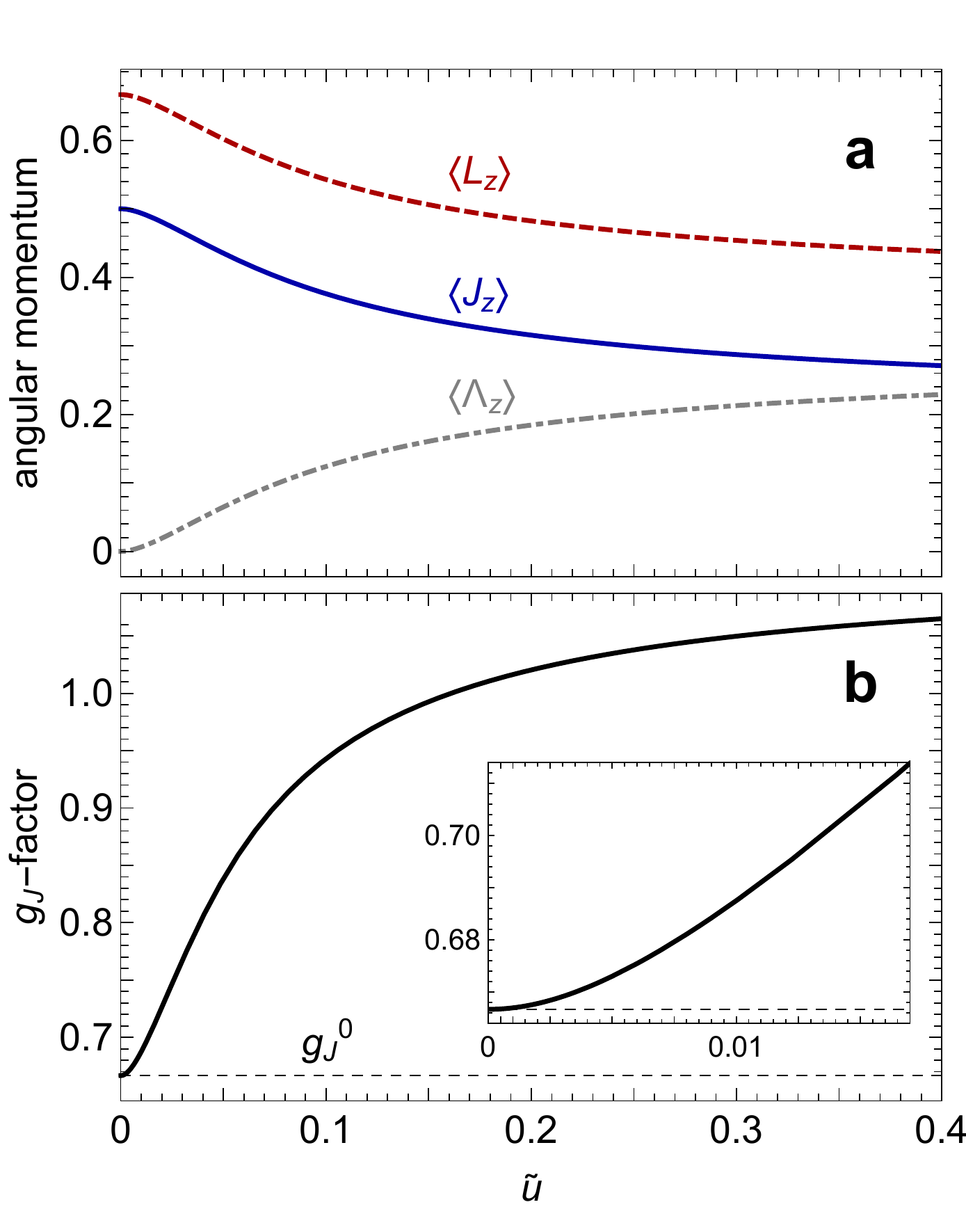}
\caption{
{\bf Static effects of phonon dressing.} {\bf a.} Quenching of orbital angular momentum by phonons. Different components of angular momentum as  a function of the dimensionless electron--phonon coupling strength, $\tilde{u}$. Due to the coupling, electronic orbital angular momentum is reduced and phonon angular momentum emerges, while the total angular momentum, $\mathsf{J}_z=J_z + \Lambda_z$, is conserved. {\bf b.} Renormalization of the electron $g$-factor as a function of the dimensionless electron--phonon coupling strength. In the perturbative regime (inset) the dependence on the coupling strength is quadratic, which can be understood as phononic Lamb shift\cite{SchmidtLem15}.
\label{fig2}}
\end{figure}

\section{Static effects of phonon dressing}
We first illustrate the appearance of non-perturbative contributions in the static case by evaluating the effect of phonon dressing on different components of angular momentum, $I_z=\langle \psi_{\mathsf{JM_J}}| \hat{I}_z|\psi_{\mathsf{JM_J}}\rangle$, where $I=L,J,\Lambda$. In the variational calculation, the electronic Hamiltonian~\eqref{eq:He} is controlled by a single parameter, which we set to $\xi/J_\text{H}=0.1$. Here we focus on the  case with $N=1$ electron, for which the configuration of the bare impurity is given by $L=1, S= J =1/2$. Furthermore, we consider the state with $M_J=J$, which is the ground state in the presence of a static magnetic field, $B_0<0$. Fig.~\ref{fig2}a shows different components of angular momentum as a function of the dimensionless electron--phonon coupling strength, $\tilde{u}=(u/E_L) \sqrt{E_M/E_L}/(2\pi^2)$. Here $u$ denotes the magnitude of the interaction $U_\lambda(k)$, $E_M=\hbar^2/(2Ma_0^2)$, with $M$ is the atomic mass of the nuclei (using $E_L=(J_\text{H}+\xi)/2$ as the unit of energy and the lattice spacing $a_0$ as the unit of length, see Appendix~\ref{app:epstrength}). In the absence of coupling, $\langle S_z \rangle=-1/6$ and $\langle L_z \rangle=2/3$, such that $\langle J_z\rangle=M_J=1/2$. While $S_z$ remains unperturbed since $\hat H_\text{ep}$ does not depend on spin, we find a reduction of orbital angular momentum that is quite distinct from the conventional picture of orbital angular momentum quenching. Instead of static crystal fields breaking rotational symmetry, here the dynamical crystal field induced by phonons causes the reduction of $\langle \hat{L}_z\rangle$ as well as of $\langle \hat{J}_z\rangle$, while conserving the total angular momentum, $\mathsf{M_J}=\langle \hat J_z \rangle + \langle \hat \Lambda_z \rangle$. The presence of phonon angular momentum also influences the response to magnetic fields. For quasi-static fields, this is reflected by the renormalization of the electron $g$-factor, which we determine from the well-known relation \cite{AshcroftMerminBook1976}:
\begin{align}
g_{J}&=\frac{g_L+g_S}{2}+\frac{g_L-g_S}{2}\frac{\langle\hat{\mathbf{L}}^2-\hat{\mathbf{S}}^2\rangle}{\langle\hat{\mathbf{J}}^2\rangle}
\end{align}
Evaluating the second term with the variational wave functions gives the result shown in Fig.~\ref{fig2}b. In the perturbative regime (inset) this yields a quadratic dependence on $\tilde{u}$ which can be understood as the phononic analog of the Lamb shift\cite{SchmidtLem15}, where virtual phonon excitations play the role of the photon excitations of quantum electrodynamics, thereby causing angular-momentum-dependent dressing of the electronic states.  For larger coupling strengths, $g_J$ features a linear dependence until signatures of saturation are observed at intermediate coupling, $\tilde{u}\sim0.4$, where the single-phonon ansatz of Eq.~\eqref{variationalpsi} becomes less reliable. Within the quasiparticle picture, the observed enhancement of the $g$-factor is analogous to the enhancement of the moment of inertia due to the formation of molecular angulons \cite{SchmidtLem15, LemeshkoDroplets16} and to the increased electron effective mass in the polaron problem.

\begin{figure}[h!]
\includegraphics[width=0.86\columnwidth]{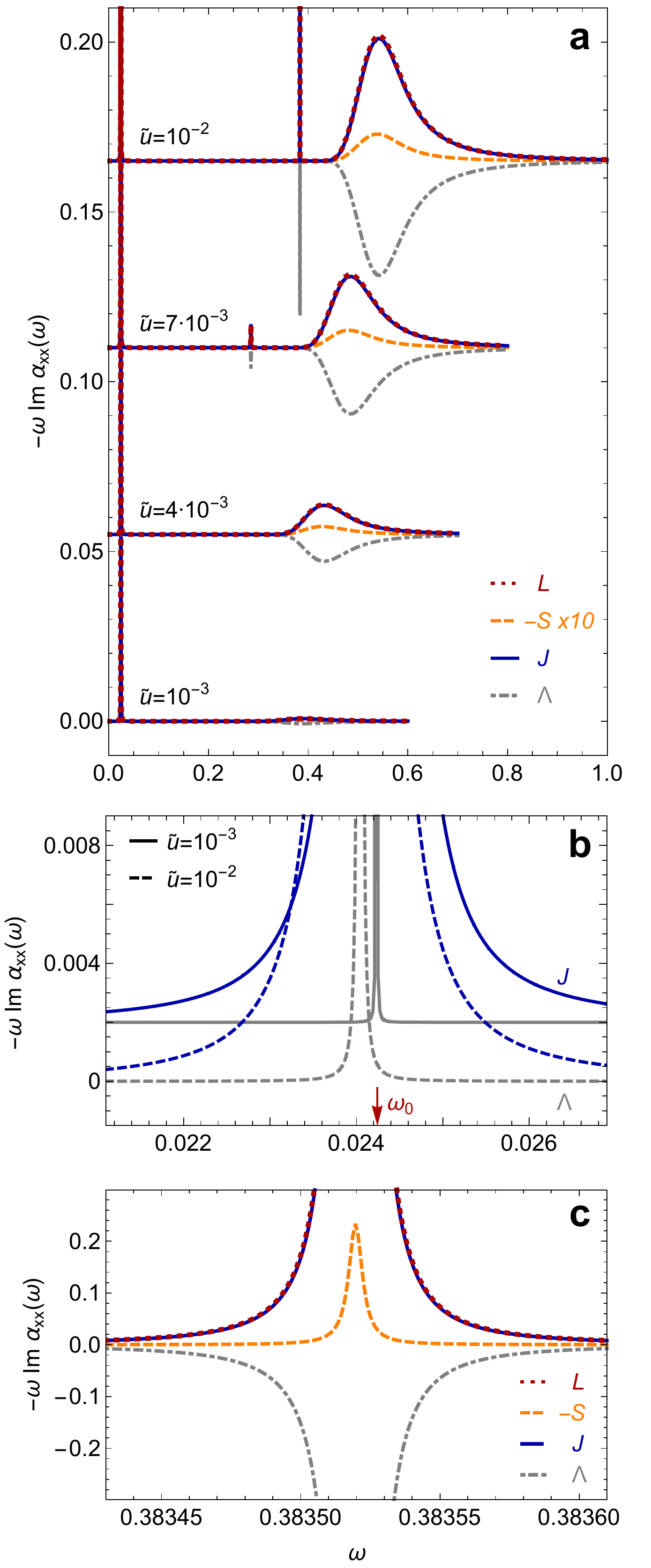}
\caption{
 {\bf Dynamical effects of phonon dressing.} {\bf a.}~Magnetic susceptibililty for various electron--phonon coupling strengths, $\tilde{u}$. Besides the low-frequency electron spin resonance (ESR) peak, a broad second spectral feature due to phonon dressing is observed.  By increasing $\tilde{u}$ gradually, a second sharp quasiparticle peak appears due to non-perturbative effects. {\bf b.}~Zoom-in in the vicinity of the ESR peak, demonstrating that phonon dressing causes a slight shift of the ESR peak to lower frequencies. {\bf c.}~Zoom-in in the vicinity of the sharp quasiparticle peak at $\tilde{u}=10^{-2}$. Exactly at the peak, the phonon susceptibility dips  due to angular momentum transfer. 
\label{linresp}}
\end{figure}

\section{Dynamic effects of phonon dressing}
Next we reveal the importance of non-perturbative effects in the dynamical response,by computing the linear response to an additional time-dependent magnetic field, $\mathbf{B}(t)=(\mathbf{B}e^{-\text{i}\omega t+\varepsilon t}+\mathbf{B}^*e^{\text{i}\omega t+\varepsilon t})/2$, $|\mathbf{B}|\ll B_0$. By determining the time-dependent changes of the variational parameters to linear order in $\mathbf{B}(t)$ and by using the general relation,
\begin{align}
\delta I_i(t)=\frac{1}{2}\sum_j\Big[\alpha^{(I)}_{ij}(\omega)B_j e^{-\text{i}\omega t+\varepsilon t} + \alpha^{(I)}_{ij}(-\omega)B^*_j e^{\text{i}\omega t+\varepsilon t}\Big]\nonumber,
\end{align}
where $\delta I_i(t)=\langle\hat{I_i}(t)\rangle-\langle\hat{I}_i\rangle_0$, $i=x,y,z$, we can derive closed-form expressions for the magnetic susceptibilities, $\alpha^{(I)}_{ij}(\omega)$, see Appendix~\ref{a3}. In Fig.~\ref{linresp} we plot $-\omega \text{Im}\alpha^I(\omega)_{xx}$ as a function of $\omega$ (in units of $E_L$), for the configuration $N=5$, $L=1, S= J =1/2, M_J=-1/2$, with electronic parameters $\xi/J_\text{H}=0.1$, $\mu_\text{B}B_0/J_\text{H}=0.02$, and various coupling strength $\tilde{u}$. For the smallest value of $\tilde{u}=10^{-3}$, the spectrum consists of a sharp peak close to the electron spin resonance (ESR) of a free atom. Furthermore, an additional broad spectral feature appears at higher energies, which is associated with incoherent phonon scattering. At low (high) frequencies, the response for $L,-S,J$ and $\Lambda$ have the same (opposite) sign, where the minus sign in $S$ comes from the fact that in the ground state $S$ is antiparallel to both $L$ and $J$. Fig.~\ref{linresp}b shows that the ESR peak width for $\Lambda$ is much narrower than that for $J$,  and increases only slightly with increasing the electron--phonon coupling strength. Hence, at these frequencies the phonons are damped much weaker than the electron spin and orbital angular momentum. This is consistent with the interpretation that the broadening of the ESR peak for  $J$ is due to the dressing with phonons in either ground or excited states, while for $\Lambda$ the decay is only possible when the dressing of distinct $M_J$ levels is different. The red arrow shows the position of the ESR peak of the free atom. We note that phonon dressing causes a shift of the ESR peak to lower frequencies, which corresponds to a reduction of the effective $g$-factor, in contrast to what is observed for the static $g$-factor in Fig.~\ref{fig2}. The static and dynamical $g$-factors are indeed two different quantities. While both can be derived from the magnetic susceptibility, the static $g_J\sim \alpha_{zz}(\omega=0)$, while the ESR peak follows from the pole of $\alpha_{xx}(\omega)$ at $\omega\neq0$. Similar differences between static and dynamical electron $g$-factors occur in the Fermi-liquid theory \cite{PlatzmanWolff1973}.

In addition to a shift of the ESR peak, Fig.~\ref{linresp}a shows that upon increasing the electron--phonon coupling strength the incoherent part moves towards higher frequencies.  Moreover, a second sharp peak gradually emerges in between the ESR peak and the incoherent part, which is shown in Figure~\ref{linresp}c for $\tilde{u}=10^{-2}$. Both high-frequency responses have opposite sign for phonon and electron angular momentum, demonstrating that magnetic fields at these frequencies induce transfer of angular momentum from electronic to lattice degrees of freedom. The second sharp peak can be identified as an additional quasiparticle peak, i.e.\ a metastable excited state of the atom dressed by additional phonons carrying angular momentum. This is reminiscent to the effect observed in conventional polaron physics. It is known as the `relaxed excited state' in the Fr\"olich model \cite{Devreese15} and as the `excited phonon-polaron bound state' in the Holstein model \cite{GogolinPSSB1982,BoncaPRB1999}, which arise at intermediate and strong coupling. Interestingly, in the present case this non-perturbative effect emerges already at weak coupling, $\tilde{u}\ll1$. The reason is that in our model the electron couples to low-energy acoustic phonons with a linear dispersion $\omega=ck$, rather than to gapped optical phonons.

The presence of additional peaks is rooted in the poles of the susceptibility, which involves additional self-consistent solutions, $E'$, to the equation $E'=E_\mathsf{JM_J} - \Sigma_\mathsf{JM_J}(E')$, where $E'>E$, $E$ being the ground-state energy. Due to the Kramers-Kronig relations, additional sharp peaks can only occur for frequencies $\omega>E'_*-E$, where $E'_*$ is defined by $\text{Max}\left[\text{Im}\,\Sigma_{\mathsf{JM_J}} (E')\right]$, since in this range $-\text{Re}\left[\Sigma_\mathsf{JM_J}(E')\right]$ is a decreasing function. Hence, there is a threshold value, $\tilde{u}_* \approx 0.0033$ for the parameters used, below which no additional metastable states can occur. Above the threshold, $\tilde{u}>\tilde{u}_*$, we find an approximately linear dependence of the position of the metastable state on the coupling strength. Such a behavior is governed by the linear dependence of the ground-state energy on $\tilde{u}$ in this regime. In Appendix~\ref{a3} we explicitly confirm this analysis numerically for various coupling strengths. We emphasise that the appearance of such additional peaks is rooted in the general Fano-type shape of $\text{Re}\left[\Sigma_\mathsf{JM_J}(E)\right]$, and should therefore be qualitatively independent of the particular approximation used to calculate the self-energy.
 
Furthermore, we obtained that upon changing the spin-orbit coupling strength, the width of the peaks changes but their position is hardly affected. These non-perturbative effects have important consequences for the dynamics. In particular, at optical frequencies, coupling of spins with a magnetic field is usually considered negligible due to the small magnitude of the magnetic component of an electromagnetic wave compared to its electric component, and the absence of magnetic dipole transitions. Our model, however, reveals that even for $u\ll J_\text{H}$, an additional resonance emerges at an energy scale of $\omega\sim J_\text{H}/5$ due to non-perturbative electron--phonon interactions. In the presence of such resonances, a magnetic field can induce transfer of angular momentum between spin and lattice degrees of freedom at ultrafast, femtosecond timescales. 

\section{Conclusions}
The results presented here demonstrate that the problem of describing the quantum dynamics of angular momentum transfer in condensed matter systems with multi-orbital atoms can be greatly simplified by casting it in terms of the angulon quasiparticles. This reformulation is achieved by deriving the electron--phonon interaction in a rotationally invariant form and using the Hubbard operators to keep track of the total angular momentum of electrons. We find that the effect of dressing of electron orbital angular momentum with phonon angular momentum leads to qualitatively new, non-perturbative high-frequency effects that should be observable in electron spin resonance experiments at THz and optical frequencies. Promising systems for experimental confirmation are paramagnetic CoO and FeO systems and nonmagnetic oxides containing orbitally degenerate impurity atoms, which, analogously to the model system studied here, contain partially filled degenerate $t_{2g}$ orbitals. While here we focused on local angular momentum transfer, which is highly relevant to understanding the fastest possible timescale for angular momentum transfer, the angulon can be used as a building block of models taking into account non-local transfer terms. Furthermore, the theory can be extended to include static crystal fields and magnetic ordering, which would pave the way to a deeper understanding of lattice dynamics during ultrafast demagnetization \cite{JalPRB17,BonettiPRL16,ReidNCOMM2018,Dornes2019}. This can potentially resolve the long-lasting debate as to whether the angular momentum transfer during ultrafast demagnetization is local or non-local and ultimately reveal the fastest possible timescale of the Einstein-de Haas and Barnett effects.


%

\section{Acknowledgements}
We acknowledge discussions with E. Yakaboylu. J.H.M. acknowledges support by the Nederlandse Organisatie voor Wetenschappelijk Onderzoek (NWO) by a VENI grant, and is part of the Shell-NWO/FOM-initiative `Computational sciences for energy research' of Shell and Chemical Sciences, Earth and Life Sciences, Physical Sciences, FOM and STW. M.I.K. acknowledges support by the European Research Council (ERC) Advanced Grant No. 338957 (FEMTO/NANO). M.L. acknowledges support from the Austrian Science Fund (FWF), under project No. P29902-N27. 

\onecolumngrid
\appendix
\begin{small}
\section{Derivation of the electron-phonon coupling Hamiltonian}\label{a1}
In this Appendix we provide details on the derivation of the rotationally invariant electron-phonon coupling Hamiltonian. In particular, we discuss the derivation of the local electron-phonon Hamiltonian, the integration over electronic and nuclear positions to derive the allowed terms respecting rotational invariance, introduce the Hubbard operators and obtain the dimensionless electron-phonon coupling strength.

\subsection{Local electron-phonon coupling}
Starting from Eq.~\eqref{eq:Hepgen}, the local electron--phonon coupling is derived by first expanding $\hat{\Psi}^\dagger(\mathbf{x})=\sum_{j}\hat{\psi}^\dagger_j(\mathbf{x}-\mathbf{r}_j)$. Inserting an identity for the nuclear density operator, $\hat{\Phi}^\dagger(\mathbf{r})\hat{\Phi}(\mathbf{r})=\sum_i\delta(\mathbf{r}-\mathbf{r}_i)$, and neglecting electron hopping between different nuclei, we have 
\begin{align}
\hat{H}_\text{ep}^\text{loc}=\sum_{ij}\int \!\!d\mathbf{x}\,\,\hat{\psi}_j^\dagger(\mathbf{x})\,V(\mathbf{x},\mathbf{r}_{ij})\,\hat{\psi}_j(\mathbf{x}).
\end{align}

\subsection{Integration over electronic coordinates}
To exploit rotational invariance of $V(\mathbf{x},\mathbf{r})=V(|\mathbf{x}-\mathbf{r}|)$, it is convenient to expand in spherical harmonics:
\begin{align}\label{Vscalar}
V(|\mathbf{x}-\mathbf{r}|)=\sum_{lm}V_l(x,r)Y^*_{lm}(\Omega_x)Y_{lm}(\Omega_r).
\end{align}
Inserting a complete set of atomic orbitals,
\begin{align}
\hat{\psi}^\dagger_j(\mathbf{x})=\sum_{\lambda\mu,\sigma}\rho_{\nu\lambda}(x)Y^*_{\lambda\mu}(\Omega_x)\chi^\dagger_{\sigma}\, \hat{c}^\dagger_{j,\lambda\mu\sigma},
\end{align}
where $\nu$ is the principal quantum number, $\lambda$ and $\mu$ are the quantum numbers for the orbital angular momentum and its projection, respectively, and $\sigma$ is the spin projection, we obtain:
\begin{align}
\label{HepLoc}
\hat{H}_\text{ep}^\text{loc}&=\sum_{ij,\sigma}\sum_{\lambda_1\mu_1}\sum_{\lambda_2\mu_{2}}\hat{c}^\dagger_{j,\lambda_1\mu_1\sigma}
\hat{c}_{j,\lambda_2\mu_2\sigma}\,
\int \!\!d\mathbf{x}\,\,\rho_{\nu\lambda_1}(x)Y^*_{\lambda_1\mu_1}(\Omega_x)\,V(\mathbf{x},\mathbf{r}_{ij})\,\rho_{\nu\lambda_2}(x)Y_{\lambda_2\mu_2}(\Omega_x),
\end{align} 
where we used that $V$ does not depend on spin. The integral in~\eqref{HepLoc} involves a radial part and an angular integral over three spherical harmonics:
\begin{align}
&\sum_{lm}\left[\int x^2dx \rho_{\nu\lambda_1}(x)\rho_{\nu\lambda_2}(x)V_l(x,r_{ij})\right]
\left[\int d\Omega_xY^*_{\lambda_1\mu_1}(\Omega_x)\,Y^*_{lm}(\Omega_x)
\,Y_{\lambda_2\mu_2}(\Omega_x)\right]
Y_{lm}(\Omega_{r_{ij}})\nonumber\\
&=\sum_{lm} g_{\lambda_1\lambda_2,l}(r_{ij})\,(-1)^mA^{\lambda_1\mu_1}_{l-m,\lambda_2\mu_2}Y_{lm}(\Omega_{r_{ij}}).
\end{align} 
Here {$g_{\lambda_1\lambda_2,l}(r_{ij})\equiv\int {dx~x^2} \rho_{\nu\lambda_1}(x)\rho_{\nu\lambda_2}(x)V_l(x,r_{ij})$} and the integration over spherical coordinates yields~\cite{VarshalovichAngMom}:
\begin{align}
A^{\lambda_1\mu_1}_{lm,\lambda_2\mu_2}=\sqrt{\frac{(2l+1)(2\lambda_2+1)}{4\pi(2\lambda_1+1)}}C^{\lambda_10}_{l0,\lambda_20}C^{\lambda_1\mu_1}_{lm,\lambda_2\mu_2},
\end{align}
where $C^{l_1m_1}_{l_2m_2,l_3m_3}$ are Clebsch-Gordon coefficients.

\subsection{Integration over nuclear coordinates}
For further derivations, we write the $\hat{H}_\text{ep}^\text{loc}$ in the form
\begin{align}
\hat{H}^{\text{loc}}_\text{ep}&=\sum_{\lambda_1\mu_1 \atop \lambda_2\mu_2}\sum_{j,\sigma}\hat{c}^\dagger_{j,\lambda_1\mu_1\sigma}
\hat{c}_{j,\lambda_2\mu_2\sigma}\frac{1}{2} \sum_{i,lm}\left[g_{\lambda_1\lambda_2,l}(r_{ij})\,Y^*_{lm}(\Omega_{ij})\,A^{\lambda_1\mu_1}_{lm,\lambda_1\mu_2}+g_{\lambda_1\lambda_2,l}(r_{ij})\,Y_{lm}(\Omega_{ij})\,(-1)^mA^{\lambda_1\mu_1}_{l-m,\lambda_1\mu_2}\right].
\end{align}
We aim to describe phonons that account for the collective dynamics of the nuclear subsystem at small deviations, $\mathbf{u}(\mathbf{r}_i)=\mathbf{r}'_i-\mathbf{r}_i$, from the equilibrium positions, $\mathbf{r}_i$. For convenience, we take the continuum limit for the nuclear coordinates $\mathbf{r}_i$ and focus on the coupling to a single atom ($j=0$). The dependence on the nuclear coordinates $\mathbf{r}$ in $H^{\text{loc}}_\text{ep}$ is then conveniently described in reciprocal space 
\begin{align}\label{F}
F^{\lambda_1\lambda_2}_{lm}(\mathbf{r}')=g_{\lambda_1\lambda_2,l}(r)Y_{lm}(\Omega_{r}) = \sum_\mathbf{k}f^{\lambda_1\lambda_2}_{lm}(\mathbf{k})e^{\text{i}\mathbf{k}\cdot\mathbf{r}}\approx F^{\lambda_1\lambda_2}_{lm}(\mathbf{r}) + \mathbf{u}(\mathbf{r})\cdot\nabla_\mathbf{r}F^{\lambda_1\lambda_2}_{lm}(\mathbf{r})
\end{align}
The term $F^{\lambda_1\lambda_2}_{lm}(\mathbf{r})$ is assumed to vanish, since it gives rise to static crystal field terms that are absent in an isotropic elastic environment. The gradient is calculated from the Fourier series:
\begin{align}
\nabla_\mathbf{r}F^{\lambda_1\lambda_2}_{lm}(\mathbf{r}) = \sum_\mathbf{k}f^{\lambda_1\lambda_2}_{lm}(\mathbf{k})\,\text{i}\mathbf{k}\,e^{\text{i}\mathbf{k}\cdot\mathbf{r}}=\frac{1}{V}\sum_\mathbf{k}\,\text{i}^{-l}\,G^{\lambda_1\lambda_2}_{l}(k)\,Y_{lm}(\Omega_k)\,\text{i}\mathbf{k}\,e^{\text{i}\mathbf{k}\cdot\mathbf{r}},\label{gradF}
\end{align}
where $V$ is the total volume of system and $f^{\lambda_1\lambda_2}_{lm}(\mathbf{k})$ is evaluated using the inversion formula and expansion of plane-waves in spherical coordinates, from which it follows that
\begin{align}
G^{\lambda_1\lambda_2}_l(k) &={4\pi}\!\!\int \!\!r^2\,dr\,g_{\lambda_1\lambda_2,l}(r)j_{l}(kr),
\end{align}
where $j_l(x)$ is the spherical Bessel function. For an isotropic elastic solid, the displacements are written in terms of phonon creation and annihilation operators as follows~\cite{KittelBook1963}:
\begin{align}\label{displacements}
\mathbf{u}(\mathbf{r}) = \mathbf{u}^\dagger(\mathbf{r}) = \frac{1}{n}\sum_{\mathbf{k}s}(2M\omega_{ks})^{-1/2}\mathbf{e}_{s}(\mathbf{k})\left[\hat{b}_{\mathbf{k}s}e^{\text{i}\mathbf{k}\cdot\mathbf{r}}+\hat{b}^\dagger_{\mathbf{k}s}e^{-\text{i}\mathbf{k}\cdot\mathbf{r}}\right],
\end{align}
where $M$ is the nuclear mass, $n$ is the number of nuclei, and $s=1,2,3$ is the polarization index. The polarization vectors, $\mathbf{e}_{s}(\mathbf{k})$, are defined by the relations $\mathbf{e}_{1}(\mathbf{k})={\mathbf{k}}/{k}$ and $\mathbf{e}_{2,3}(\mathbf{k})\cdot\mathbf{k}=0$ for longitudinal and transverse phonons, respectively. Hence, from evaluating the scalar product in Eq.~\eqref{F} using Eq.~\eqref{displacements} and Eq.~\eqref{gradF}, we obtain that only longitudinal phonons contribute. We drop the label $s=1$ below and obtain:
\begin{align}
\frac{1}{V_r}\!\int\!d\mathbf{r}\,\mathbf{u}(\mathbf{r})\cdot\nabla_\mathbf{r}F^{\lambda_1\lambda_2}_{lm}(\mathbf{r})&=\frac{1}{V}\sum_\mathbf{k}(2M\omega_{k})^{-1/2}G^{\lambda_1\lambda_2}_l(k)\,(\text{i}k)\,\text{i}^{-l}\,\left[-Y_{lm}(\Omega_{-k})\,\hat{b}_{\mathbf{-k}}+Y_{lm}(\Omega_k)\,\hat{b}^\dagger_{\mathbf{k}}\right].
\end{align}
Here $V_r=V/n$ is the volume of the unit cell and we used $\int\!d\mathbf{r} e^{\text{i}\mathbf{k}\mathbf{r}}=(2\pi)^3\delta({\mathbf{k}})$. We are still left with the dependence on angles, $\Omega_k$, which can be removed by transforming to spherical phonon operators using the definition \cite{LemSchmidtChapter}:
\begin{align}
\hat{b}^\dagger_{\mathbf{k}}=\frac{(2\pi)^{3/2}}{k}\sum_{\lambda\mu}\hat{b}_{k\lambda\mu}^\dagger \text{i}^{\lambda}Y^*_{\lambda\mu}(\Omega_k). \label{bdagtransf}
\end{align}
Using
\begin{align}
\sum_\mathbf{k}=\frac{1}{V_k(2\pi)^3}\!\!\int \!\!dk\,k^2\!\!\int\!\! d\Omega_k,
\end{align}
we can integrate over angles in $k$-space yielding
\begin{align}
\frac{1}{V_r}\!\int\!d\mathbf{r}\,\mathbf{u}(\mathbf{r})\cdot\nabla_\mathbf{r}F^{\lambda_1\lambda_2}_{lm}(\mathbf{r})&=\sum_{k\lambda\mu} U_\lambda(k)\,\text{i}\left[-\hat{b}_{k\lambda\mu}(-1)^{\lambda}(-1)^{\mu}\delta_{\lambda l}\delta_{\mu-m}+ \hat{b}^\dagger_{k\lambda\mu}\delta_{\lambda l}\delta_{\mu m}\right],
\end{align}
where $\sum_k \equiv \int \!dk$ and 
\begin{align}\label{e:Uintegral}
U^{\lambda_1\lambda_2}_\lambda(k) = \frac{1}{(2\pi)^{3/2}}k^2(2M\omega_{k})^{-1/2}G^{\lambda_1\lambda_2}_\lambda(k).
\end{align}
Finally, we obtain:
\begin{align}
\hat{H}^{\text{loc}}_\text{ep}
&=\sum_{\lambda_1\mu_1\atop \lambda_2\mu_2}\sum_{k\lambda\mu}\sum_{j\sigma}\hat{c}^\dagger_{j\lambda_1\mu_1\sigma}\hat{c}_{j\lambda_2\mu_2\sigma}\,U^{\lambda_1\lambda_2}_\lambda(k)\frac{\text{i}}{2}\left(1+(-1)^\lambda\right)\left[-A^{\lambda_1\mu_1}_{\lambda\mu,\lambda_1\mu_2}\hat{b}_{k\lambda\mu} +  (-1)^\mu A^{\lambda_1\mu_1}_{\lambda-\mu,\lambda_1\mu_2}\hat{b}^\dagger_{k\lambda\mu}\right].
\end{align}
The factor of $(1-(-1)^\lambda)$ originates from the assumption of an isotropic elastic solid, which ensures that only even $\lambda$ contributes to the transfer of angular momentum. Hermicity, $\hat{H}^{\text{loc}}_\text{ep}=\left(\hat{H}^{\text{loc}}_\text{ep}\right)^\dagger$, is easily proved using the symmetry relations for the Clebsch-Gordon coefficients, $C^{\lambda_1\mu_2}_{\lambda\mu,\lambda_1\mu_1}=(-1)^\mu C^{\lambda_1\mu_1}_{\lambda-\mu,\lambda_1\mu_2}$. 

\subsection{Hubbard operators}
Since $V(\mathbf{x},\mathbf{r})$ does not depend on spin, phonons only change the total orbital angular momentum $L$ of the electrons. This is made explicit by transforming from single-electron operators to many-electron $\hat X$-operators \cite{IrkhinPSSB1994} (also known as Hubbard operators \cite{HubbardPRSA1965}),  $\hat X(\Gamma,\Gamma')=|\Gamma\rangle\langle \Gamma'|$. For a general operator acting on a single site $i$ we have 
\begin{equation}
\hat{O}_i=\sum_{\Gamma,\Gamma'}\langle\Gamma|\hat{O}_i|\Gamma'\rangle \hat X_i(\Gamma,\Gamma'),\quad  \hat X_i(\Gamma,\Gamma')=|i\Gamma\rangle\langle i \Gamma'|.
\end{equation}
Here $\Gamma=NLMS\mathit{\Sigma}\alpha$ are the quantum numbers of many-electron states, where $N$ denotes the total number of electrons, $L,S$ are total orbit and spin quantum numbers with projections $M,\mathit{\Sigma}$, and $\alpha$ is the Racah seniority quantum number. For the single-electron creation operator the matrix element reads~\cite{IrkhinPSSB1994,IrkhinBook2000}:
\begin{align}\label{cdaggerX}
\langle\Gamma_N|\hat{c}^\dagger_{i\lambda\mu\sigma}|\Gamma_{N-1}\rangle=N^{1/2}G^{\Gamma_N}_{\Gamma_{N-1}}C^{\Gamma_N}_{\Gamma_{N-1},\lambda\mu\sigma},
\end{align}
where $G^{\Gamma_N}_{\Gamma_{N-1}}=G^{L_NS_N}_{L_{N-1}S_{N-1}}$ is the coefficient of fractional parentage~\cite{RudzikasBook2007}, and $C^{\Gamma_N}_{\Gamma_{N-1},\gamma}$ is expressed through the Clebsch-Gordon coefficients as
\begin{align}
C^{\Gamma_N}_{\Gamma_{N-1},\gamma}=C^{L_NM_N}_{L_{N-1}M_{N-1},lm}C^{S_N\mathit{\Sigma}_N}_{S_{N-1}\mathit{\Sigma}_{N-1},s\sigma},
\end{align}
with $s=1/2$. Using Eq.~\eqref{cdaggerX} we obtain
\begin{align}\label{cdaggercX}
\langle\Gamma_N|\hat{c}^\dagger_{i\lambda_1\mu_1\sigma_1}\hat{c}_{i\lambda_2\mu_2\sigma_2}|\Gamma'_{N}\rangle=N \sum_{\Gamma''_{N-1}}G^{\Gamma_N}_{\Gamma''_{N-1}}C^{\Gamma_N}_{\Gamma''_{N-1},\lambda_1\mu_1\sigma_1}G^{\Gamma'_N}_{\Gamma''_{N-1}}C^{\Gamma'_N}_{\Gamma''_{N-1},\lambda_2\mu_2\sigma_2}.
\end{align}
In the electron-phonon coupling only the summation over single-electron operators with the same spin $\sigma_1=\sigma_2$ enters,
\begin{align}
\sum_\sigma\hat{c}^\dagger_{i\lambda_1\mu_1\sigma}\hat{c}_{i\lambda_2\mu_2\sigma} &=
N \sum_{\Gamma''_{N-1}}G^{\Gamma_N}_{\Gamma''_{N-1}}C^{\Gamma_N}_{\Gamma''_{N-1},\lambda_1\mu_1\sigma}G^{\Gamma'_N}_{\Gamma''_{N-1}}C^{\Gamma'_N}_{\Gamma''_{N-1},\lambda_2\mu_2\sigma}\hat X_i(\Gamma_N,\Gamma'_N),
\end{align}
which ensures that only states $S'=S$, $\mathit{\Sigma}_N'=\mathit{\Sigma}_N$ contribute, as follows from summation over both $\mathit{\Sigma}_N''$ and $\sigma$ and by using the unitarity relation for the Clebsch-Gordan coefficients. For example, for a three-orbital atom, we have $\lambda_1=\lambda_2$ and the seniority quantum number can be omitted. In this case we obtain we obtain the following coupling term:
\begin{align}
W^{L_NL_N'S_N}_{M_NM_N'\mu_1\mu_2}&=\sum_{L''_{N-1}S''_{N-1}} G^{L_NS_N}_{L''_{N-1}S''_{N-1}}G^{L'_NS_N}_{L''_{N-1}S''_{N-1}}\sum_{M''_{N-1}}C^{L_NM_N}_{L''_{N-1}M''_{N-1},\lambda_1\mu_1} C^{L'_NM'_N}_{L''_{N-1}M''_{N-1},\lambda_1\mu_2},
\end{align}
yielding
\begin{align}
\sum_\sigma\hat{c}^\dagger_{i\lambda_1\mu_1\sigma}\hat{c}_{i\lambda_1\mu_2\sigma} &= \sum_{NS_N\mathit{\Sigma}_N}\sum_{L_NL'_NM_NM_N'}N\,W^{L_NL'_NS_N}_{M_NM_N'\mu_1\mu_2}\,X_i(NL_NM_NS_N\mathit{\Sigma}_N,NL'_NM'_NS_N\mathit{\Sigma}_N)
\end{align}
Note that it follows from the symmetry of the Clebsch-Gordan coefficients that only $M'_N=M_N-\mu_1+\mu_2$ remains in the summation.

\subsection{Electron-phonon coupling strength}\label{app:epstrength}
For numerical calculations we need to evaluate the radial integrals in $U_\lambda(k)=U^{\lambda_1\lambda_1}_\lambda(k)$ (see Eq.~\eqref{e:Uintegral}) for which we use Gaussian form factors, $g_\lambda(r)=\frac{u_\lambda}{(2\pi)^{3/2}}e^{-r^2/(2r_\lambda^2)}$, where $u_\lambda$ parametrizes the strength of the electron-phonon coupling. Introducing dimensionless units, with $E_L=(J_\text{H}+\xi)/2$ being the unit of energy and the lattice spacing $a_0$ being the unit of length, we can write
\begin{align}
\tilde{U}_\lambda(\tilde{k}) &= {{\tilde{u}}_\lambda}\,\frac{\tilde{k}^{3/2}}{\tilde{c}^{1/2}}\int_0^\infty \! \!\tilde{r}^2\,d\tilde{r}\,e^{-\tilde{r}^2/(2\tilde{r}^2_\lambda)}j_{\lambda}(\tilde{k}\tilde{r})
\end{align}
We use $\tilde{r}_\lambda=1$ to characterize the interaction range. The interaction strength is parametrized by $\tilde{u}_0=\tilde{u}$, $\tilde{u}_2=0.5\tilde{u}$. For the dimensionless electron-phonon coupling strength we obtain $\tilde{u}=(u/E_L) \sqrt{E_M/E_L}/(2\pi^2)$, where $E_M=\hbar^2/2Ma_0^2$. For transition metal atoms $M\sim100\cdot10^{-27}$~kg, $a_0\sim2\AA$, $E_L\sim0.5$~eV, we have $E_M/E_L\sim0.1$, which ensures that $\tilde{u}\ll1$ even if $(u/E_L)\sim1$. For the dimensionless speed of sound we use $\tilde{c}=0.05$, consistent with $c\sim 3-6 \cdot 10^3$~m/s for typical solid-state systems.


\section{Variational solution for the static case}\label{a2}
\subsection{Non-perturbative self-energy}
Here we discuss the derivation of the variational solution in more detail, providing explicit expressions for the matrix elements that enter the final result. For the static case, we deal with the Hamiltonian
\begin{align}
\hat H = \hat H_\text{e}^\text{C}+ \hat H^{LS}_\text{e} + \hat H_\text{p} + \hat H_\text{ep}^{\text{loc}} + \hat H_Z,
\end{align}
where $\hat H_Z=\mu_\text{B}B_0\left(g_L\hat{L}^z+g_S\hat{S}^z\right)$. Owing to the presence of spin-orbit coupling, only $\Gamma=LSJM_J$ are good quantum numbers, where $\hat{\mathbf{J}}=\hat{\mathbf{L}}+\hat{\mathbf{S}}$ with projection $M_J$ and for brevity we omit the label $N$. In addition, since $\hat H_\text{ep}$ couples directly only to orbital momentum $L$, we choose the variational wavefunctions as follows:
\begin{align}
|\psi_\mathsf{JM_J}\rangle & = Z^{1/2}_\mathsf{JM_J}|LSJM_J\rangle|0\rangle + \sum_{k\lambda\mu}\sum_{lm}\beta^\mathsf{JM_J}_{k\lambda l}\,\sum_{M\mathit{\Sigma}} C^{JM_J}_{LM,S\mathit{\Sigma}} C^{LM}_{lm,\lambda\mu} \hat{b}^\dagger_{k\lambda\mu}|0\rangle|lmS\mathit{\Sigma}\rangle= |\psi_1\rangle + |\psi_2\rangle,
 \end{align}
where we used that $|LSJM_J\rangle = \sum_{M\mathit{\Sigma}} C^{JM_J}_{LM,S\mathit{\Sigma}}|LMS\mathit{\Sigma}\rangle
$. $Z^{1/2}_{JM_J}$ and $\beta^{(JM_J)}_{k\lambda l}$ are variational parameters to be determined from minimizing $ E= \langle\psi |\hat H|\psi\rangle / \langle\psi |\psi\rangle$. This is equivalent to minimizing $F=\langle\psi |\hat H-E|\psi\rangle$ and the following terms enter
\begin{align}
\langle\psi_1|\hat H^\text{C}_\text{e}+\hat H^{LS}_\text{e}+\hat H_Z+\hat H_p + \hat H_Z|\psi_1\rangle &= E_\mathsf{JM_J}|Z_\mathsf{JM_J}^{1/2}|^2 \\
\langle\psi_2|\hat H^\text{C}_\text{e}+\hat H^{LS}_\text{e}+\hat H_Z+\hat H_p + \hat H_Z|\psi_2\rangle &= \sum_{k\lambda l} (E^\mathsf{JM_J}_{\lambda l}+\omega_k)|\beta^\mathsf{JM_J}_{k\lambda l}|^2  \\
E_\mathsf{JM_J} &=E_{NS} + E_LL(L+1) + E_J J(J+1) + E_Z M^z_{M_J}\\
E^\mathsf{JM_J}_{\lambda l} &=E_{NS} + E_Ll(l+1) + E_J P_{\lambda l} + E_Z m^z_{\lambda l}
\end{align}
Here $E_L=-J_\text{H}-\xi/2$, $E_J=\xi/2$, and $E_{NS}$ is the energy term depending on $N$ and $S$ which remains constant in the variational solution. Furthermore, we have defined
\begin{align}
M^z_{M_J}=\sum_{M\mathit{\Sigma}}\left(C^{JM_J}_{LM,S\mathit{\Sigma}}\right)^2\left[g_LM+g_S\mathit{\Sigma}\right],\qquad m^z_{\lambda l}=\sum_{M\mathit{\Sigma}}\sum_{m\mu}\left(C^{JM_J}_{LM,S\mathit{\Sigma}}\right)^2\left(C^{LM}_{lm,\lambda\mu}\right)^2\left[g_Lm+g_S\mathit{\Sigma}\right]
\end{align}
as well as the bare spin-orbit coupling terms in the atomic state with phonons excited,
\begin{align}
P_{\lambda l}&=\sum_{MM'\mathit{\Sigma}}\sum_{m,jm_j}C^{JM_J}_{LM,S\mathit{\Sigma}}C^{JM_J}_{LM'\!,S\bar{\mathit{\Sigma}}}\,C^{LM}_{lm,\lambda(M-m)}C^{LM'}_{l\bar{m},\lambda(M-m)}\,C^{jm_j}_{lm,S\mathit{\Sigma}}C^{jm_j}_{l\bar{m},S\bar{\mathit{\Sigma}}}j(j+1),
\end{align}
where $\bar{m}=m-(M-M')$, $\bar{\mathit{\Sigma}}=\mathit{\Sigma}+(M-M')$. In addition we have
\begin{align}
\langle\psi_1|\hat H^{\text{loc}}_\text{ep}|\psi_2\rangle &= -\text{i}\,Z^{1/2*}_\mathsf{JM_J}\sum_{k\lambda l}\beta^\mathsf{JM_J}_{k\lambda l}U_\lambda(k)Q_{\lambda l},\\
Q_{\lambda l}&=\frac{1}{2}\left(1+(-1)^\lambda\right)N\sum_{\mathit{\Sigma}M}\left(C^{JM_J}_{LM,S\mathit{\Sigma}}\right)^2\,\left(\sum_{\mu_1\mu_2}A^{\lambda_1\mu_1}_{\lambda(\mu_1-\mu_2),\lambda_1\mu_2}W^{LlS}_{M\bar{M}\mu_1\mu_2}C^{LM}_{l\bar{M},\lambda(\mu_1-\mu_2)}\right), 
\end{align}
with $\bar{M}=M-\mu_1+\mu_2$ and $U_\lambda(k)=U^{\lambda_1\lambda_1}_\lambda(k)$. In deriving these expressions we have used several times the symmetry properties of the Clebsch-Gordon coefficients. Minimization gives the equations 
\begin{align}
E &= E_\mathsf{JM_J} - \Sigma_\mathsf{JM_J}(E),\quad \Sigma_\mathsf{JM_J} (E)=\sum_{k\lambda l} \frac{U_\lambda(k)^2Q_{\lambda l}^2}{E^\mathsf{JM_J}_{\lambda l}- E +\omega_k}.
\end{align}

Once $E$ is obtained, the variational parameters can be determined from the relations
\begin{align}
\frac{\beta^\mathsf{JM_J}_{k\lambda l}}{Z_\mathsf{JM_J}^{1/2}}=\frac{-\text{i}\,U_\lambda(k)\,Q_{\lambda l}}{E - E^\mathsf{JM_J}_{\lambda l}-\omega_k }
\equiv R^{\mathsf{JM_J}}_{k\lambda l}(E),\quad
|Z_\mathsf{JM_J}^{1/2}|=\left(1+\sum_{k\lambda l}|R^{\mathsf{JM_J}}_{k\lambda l}(E)|^2\right)^{-1/2}.
\end{align}

\subsection{$g$-factor renormalization}
Once the variational parameters are determined, observables can be directly evaluated. For the calculation of the $g$-factor we need to evaluate
\begin{align}
g_{J}&=\frac{g_L+g_S}{2}+\frac{g_L-g_S}{2}\frac{\langle\hat{\mathbf{L}}^2-\hat{\mathbf{S}}^2\rangle}{\langle\hat{\mathbf{J}}^2\rangle}
\end{align}
Direct substitution gives
\begin{align}
\frac{\langle\psi|\hat{\mathbf{L}}^2-\hat{\mathbf{S}}^2|\psi\rangle}{\langle\psi|\hat{\mathbf{J}}^2|\psi\rangle}=\frac{|Z_\mathsf{JM_J}^{1/2}|^2\big[L(L+1)-S(S+1)\big]+\sum_{k\lambda l}|\beta^\mathsf{JM_J}_{k\lambda l}|^2 \big[l(l+1)-S(S+1)\big]}{|Z_\mathsf{JM_J}^{1/2}|^2J(J+1) + \sum_{k\lambda l}|\beta^\mathsf{JM_J}_{k\lambda l}|^2 P_{\lambda l}}
\end{align}
For weak electron-phonon interactions we have $|{\beta^\mathsf{JM_J}_{k\lambda l}}/{Z_\mathsf{JM_J}^{1/2}}|^2=|R^\mathsf{JM_J}_{k\lambda l}|^2\ll1$ and we can write 
\begin{align}
g_{J}&\approx g_J^0 + \frac{g_L-g_S}{2}\sum_{k\lambda l}|R^\mathsf{JM_J}_{k\lambda l}|^2 \left[\frac{l(l+1)-S(S+1)}{J(J+1)} -\frac{L(L+1)-S(S+1)}{J(J+1)}\frac{P_{\lambda l}}{J(J+1)}\right],\label{geff}
\end{align}
where
\begin{align}
g_J^0=\frac{g_L+g_S}{2} + \frac{g_L-g_S}{2}\frac{L(L+1)-S(S+1)}{J(J+1)}.
\end{align}
Hence, at small coupling we expect a change of the $g$-factor that scales quadratically with the coupling strength.

\section{Variational solution for dynamical response}\label{a3}
\subsection{Linear response formulas}\label{a:linrespformulas}
To derive the equations for linear response, the variational parameters are written as 
\begin{align}
Z^{1/2}_\mathsf{JM_J}(t)&=Z^{1/2}_\mathsf{JM_J}+\delta Z^{1/2}_\mathsf{JM_J}(t) \\
\beta^{\mathsf{JM_J}}_{k\lambda\mu}(t)&=\beta^{\mathsf{JM_J}}_{k\lambda\mu}+\delta \beta^{\mathsf{JM_J}}_{k\lambda\mu}(t),
\end{align}
where 
\begin{align}
\delta Z^{1/2}_\mathsf{JM_J}(t) &= \delta Z^{1/2}_\mathsf{JM_J}(\omega)e^{-\text{i}\omega t+\varepsilon t} + \delta Z^{1/2}_\mathsf{JM_J}(-\omega)e^{\text{i}\omega t+\varepsilon t},\nonumber\\
\delta \beta^\mathsf{JM_J}_{k\lambda l}(t) &= \delta \beta^\mathsf{JM_J}_{k\lambda l}(\omega)e^{-\text{i}\omega t+\varepsilon t} + \delta \beta^\mathsf{JM_J}_{k\lambda l}(-\omega)e^{\text{i}\omega t+\varepsilon t}.\nonumber
\end{align}
and $Z^{1/2}_{\mathsf{JM_J}}$ and $\beta^{\mathsf{JM_J}}_{k\lambda\mu}$ are given by the solution of the static case. For convenience we write the dynamical contributions as
\begin{align}
\delta Z_\mathsf{JM_J}^{1/2}(\omega) &= \frac{1}{2}\mathbf{B}\cdot\boldsymbol{X}_\mathsf{JM_J}(\omega)\nonumber\\
\delta \beta^\mathsf{JM_J}_{k\lambda l}(\omega)&=\frac{1}{2}\mathbf{B}\cdot\boldsymbol{\chi}^\mathsf{JM_J}_{k\lambda l}(\omega),\nonumber
\end{align}
from which $\delta Z_\mathsf{JM_J}^{1/2}(-\omega)$ and $\delta \beta^\mathsf{JM_J}_{k\lambda l}(-\omega)$ are obtained by replacing $\omega\rightarrow -\omega$ and $\mathbf{B}\rightarrow\mathbf{B}^*$. In this notation,  evaluation of the time-dependent changes of angular momentum,  $\delta I_i(t)=\langle\hat{I_i}(t)\rangle-\langle\hat{I}_i\rangle_0$, yields the susceptibilities
\begin{align}
\alpha^{(I)}_{ij}(\omega)=&\sum_\mathsf{M_JM_J'}\bigg[I_{i}^\mathsf{M_JM_J'}\big[Z_\mathsf{JM_J}^{1/2*} {X}_{\mathsf{JM_J'},j}(\omega) + Z_\mathsf{JM_J'}^{1/2} \,{X}^*_{\mathsf{JM_J},j}(-\omega)\big]+ \sum_{k\lambda l}I_{\lambda l,i}^\mathsf{M_JM_J'}\big[\beta^{\mathsf{JM_J}*}_{k\lambda l}\,{\chi}^{\mathsf{JM_J'}}_{k\lambda l,j}(\omega) + \beta^{\mathsf{JM_J'}}_{k\lambda l}\,{\chi}^{\mathsf{JM_J}*}_{k\lambda l,j}(-\omega)\big]\bigg]\nonumber,
\end{align}
where 
\begin{align}
I_{i}^\mathsf{M_JM_J'}&=\langle LS{JM_J}|\hat{I}_i|LS{JM_J'}\rangle,\\
I_{\lambda l,i}^\mathsf{M_JM_J'}&=\!\!\sum_{MM'\mathit{\Sigma}\mathit{\Sigma}'}\!\!C^{JM_J}_{LM,S\mathit{\Sigma}}C^{JM_J'}_{LM',S\mathit{\Sigma}'}\!\!\sum_{mm'\mu\mu'}\!\!C^{LM}_{lm,\lambda\mu}C^{LM'}_{lm',\lambda\mu'}
\langle lmS\mathit{\Sigma}\lambda\mu|\hat{I}_i|lm'S\mathit{\Sigma}'\lambda\mu'\rangle,
\end{align}
are the matrix elements of angular momentum components $I=L,S,J,\Lambda$ without and with phonons excited, respectively, with $i=x,y,z$. For practical calculations we focus to the case of a non-degenerate ground state. For example, at $B_0>0$ the variational state with $\mathsf{M^0_J}=\mathsf{-J}$ has the lowest energy. Hence, in the static problem only $Z_\mathsf{JM_J^0}^{1/2}$ and $\beta_{k\lambda l}^\mathsf{JM_J^0}$ are nonzero. For numerical evaluation of $\alpha^{(I)}_{ij}(\omega)$ it is convenient to determine the contributions of ${X}_{\mathsf{JM_J},i}(\omega)$ and ${\chi}^{\mathsf{JM_J}}_{k\lambda l,i}(\omega)$ from the expressions:
\begin{align}
&X_{\mathsf{JM_J},i}(\omega) = Z^{1/2}_{\mathsf{JM_J^0}}\,\frac{M^\mathsf{M_JM_J^0}_i + \sum_{\lambda l} m_{\lambda l,i}^\mathsf{M_JM_J^0}\,\text{K1}^\mathsf{M_J}_{\lambda l}(\omega)}{\omega + \text{i}\varepsilon + \Delta E_\mathsf{JM_J} - \text{K0}^\mathsf{M_J}(\omega)},
\end{align} 
\begin{align}
\sum_k\beta_{k\lambda l}^{\mathsf{JM_J^0}*}{\chi}^{\mathsf{JM_J}}_{k\lambda l,i}(\omega)=&Z^{1/2}_\mathsf{JM_J^0}\,\text{K1}^\mathsf{M_J}_{\lambda l}(\omega){X}_{\mathsf{JM_J},i}(\omega) + |Z^{1/2}_\mathsf{JM_J^0}|^2\,\text{K2}^\mathsf{M_J}_{\lambda l}(\omega)\,{m}^\mathsf{M_JM_J^0}_{\lambda l,i},\nonumber
\end{align} 
where
\begin{align}
{M}^\mathsf{M_JM_J'}_i&=\mu_\text{B}\left(g_L{L}^\mathsf{M_JM_J'}_i+g_S{S}^\mathsf{M_JM_J'}_i\right),\\
{m}^\mathsf{M_JM_J'}_{\lambda l,i} &=\mu_\text{B}\left(g_L{L}^\mathsf{M_JM_J'}_{\lambda l,i}+g_S{S}^\mathsf{M_JM_J'}_{\lambda l,i}\right),
\end{align} 
and
\begin{align}
\text{K0}^\mathsf{M_J}(\omega)&=\sum_{k\lambda l} \frac{|U_\lambda(k)\,Q_{\lambda l}|^2}{\big(\omega + \text{i}\varepsilon + \Delta E^{\mathsf{JM_J}}_{\lambda l}-\omega_k\big)}\\
\text{K1}^\mathsf{M_J}_{\lambda l}(\omega)&=\sum_k \frac{|U_\lambda(k)\,Q_{\lambda l}|^2}{\big(\Delta E^{\mathsf{JM_J^0}}_{\lambda l}-\omega_k\big)\big(\omega + \text{i}\varepsilon + \Delta E^{\mathsf{JM_J}}_{\lambda l}-\omega_k\big)}\\
\text{K2}^\mathsf{M_J}_{\lambda l}(\omega)&=\sum_k \frac{|U_\lambda(k)\,Q_{\lambda l}|^2}{\big(\Delta E^{\mathsf{JM_J^0}}_{\lambda l}-\omega_k\big)^2\big(\omega + \text{i}\varepsilon + \Delta E^{\mathsf{JM_J}}_{\lambda l}-\omega_k\big)}
\end{align} 
with $\Delta E_\mathsf{JM_J}=E-E_\mathsf{JM_J}$ and $\Delta E^{\mathsf{JM_J}}_{\lambda l}=E-E^\mathsf{JM_J}_{\lambda l}$, where $E$ is the variational ground-state energy.  The integrals are computed numerically with $\varepsilon\ll1$ until convergence is achieved. Explicit expressions for the energy of the bare impurity, $E_\mathsf{JM_J}$, and the energy of the bare impurity with phonons excited, $E^\mathsf{JM_J}_{\lambda l}$, as well as for $Q_{\lambda l}$, are given in Appendix~\ref{a2}.

{\subsection{Emergence of high-frequency peaks}\label{app:peaks}
The emergence of high-frequency peaks in the susceptibilities can be understood by analyzing (i) the poles of $X_{\mathsf{JM_J},i}$ (Eq.~(22) of the Methods section), which involves changes of $Z_\mathsf{JM_J}^{1/2}(\omega)$ and (ii) the functions $\text{K}1^\mathsf{M_J}_{\lambda l}(\omega)$ and $\text{K}2^\mathsf{M_J}_{\lambda l}(\omega)$ (Eqs.~(27)--(28) of the Methods section), which corresponds to changes of $\beta^\mathsf{JM_J}_{k\lambda l,i}(\omega)$. Here we elaborate on the poles of $X_{\mathsf{JM_J},i}$ which give rise to additional metastable states of the quasiparticle and are determined by the equation
\begin{align}\label{polesX}
\omega + \Delta E_\mathsf{JM_J} - \text{K0}^\mathsf{M_J}(\omega)=0,
\end{align}
where $\Delta E_\mathsf{JM_J}=E-E_\mathsf{JM_J}$. Using $\omega=E'-E$, with $E$ the ground-state energy and the definition of $\text{K0}^\mathsf{M_J}(\omega)$ (see Eq. (26) of the Methods section), we find that the solution of \eqref{polesX} coincides with the solutions of $E'=E_\mathsf{JM_J}-\Sigma_{\mathsf{JM_J}}(E')$  corresponding to the angulon states at energies $E'>E$. Such additional solutions only occur for sufficiently high electron--phonon coupling strength, as we illustrate in Fig.~\ref{graphicalsolution} by plotting $E'-E_\mathsf{JM_J}$ (gray solid line), $-\text{Re}\left[\Sigma_{\mathsf{JM_J}}(E')\right]$ (blue dashed line) and $\text{Im}\left[\Sigma_{\mathsf{JM_J}}(E')\right]$ (red dotted line) as a function of $E'$ for a few different electron--phonon coupling strengths. Vertical dashed and dotted lines indicate the variational ground state energies of the angulon and of the free atom, respectively. We observe that additional sharp peaks only occur for $E'\gg E'_*$, where $E'_*$ is defined by $\text{Max}\left[\text{Im}\,\Sigma_{\mathsf{JM_J}} (E')\right]$, since in this range $-\text{Re}\,\Sigma_{\mathsf{JM_J}}(E')$ is a monotonically decreasing function and $\text{Im}\,\Sigma_{\mathsf{JM_J}} (E')$ remains small but not negligible.}

\begin{figure}[h!]
\includegraphics[width=1.0\columnwidth]{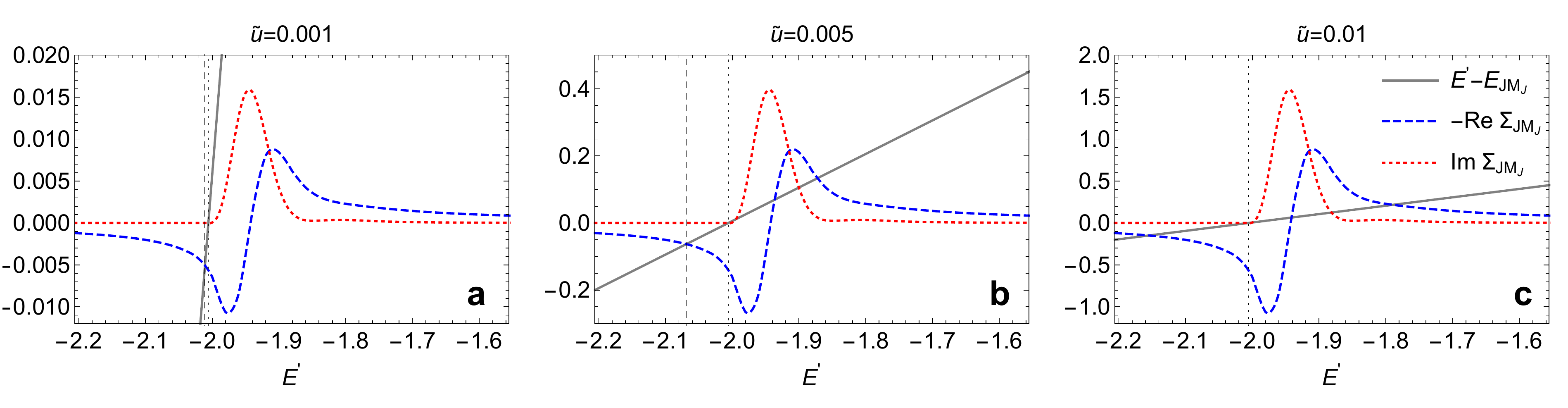}
\caption{
 {{\bf Emergence of metastable states due to the Fano-like shape of the self-energy.} Self-consistent solutions are determined by crossings of the solid grey line and the blue dashed line, which represents the real part of the self-energy, $-\text{Re}\,\Sigma_{\mathsf{JM_J}}(E')$. {\textbf{a.}} For the smallest electron--phonon coupling strength, $\tilde{u}=0.001$, only one self-consistent solution is found (vertical dashed line) at a slightly lower energy than the ground-state energy of the bare atom (vertical dotted line). {\textbf{b.}} For larger coupling strengths,  additional self-consistent solutions emerge. The metastable state corresponds to the solution with the largest energy. {\textbf{c.}} By further increasing the coupling strength, the metastable state shifts towards higher energies $E'$ where $-\text{Re}\,\Sigma_{\mathsf{JM_J}}(E')$ monotonically decreases. For comparison, the imaginary part of the self-energy is shown (red dotted line), which determines the lifetime of the metastable state. Note the different scales of the vertical axes for different coupling strengths.} 
\label{graphicalsolution}}
\end{figure}

{
In addition, in order to investigate the scaling of the high-frequency peak with electron--phonon coupling strength $\tilde{u}$, we evaluate the dependences $E(\tilde{u})$ and $E'(\tilde{u})$. The result is shown in Fig.~\ref{scaling}, by solid black ($E$) and dashed blue ($E'$) lines, from which we conclude that the change of $\omega(\tilde{u})=E'(\tilde{u})-E(\tilde{u})$ is approximately linear with $\tilde{u}$. The results shown in Fig.~\ref{graphicalsolution} and Fig.~\ref{scaling} are computed for the same parameters as Fig. 3 of the main text: $N=5$, $L=1, S= J =1/2, M_J=-1/2$, with electronic parameters $\xi/J_\text{H}=0.1$, $\mu_\text{B}B_0/J_\text{H}=0.02$.
}

\begin{figure}[h!]
\includegraphics[width=0.45\columnwidth]{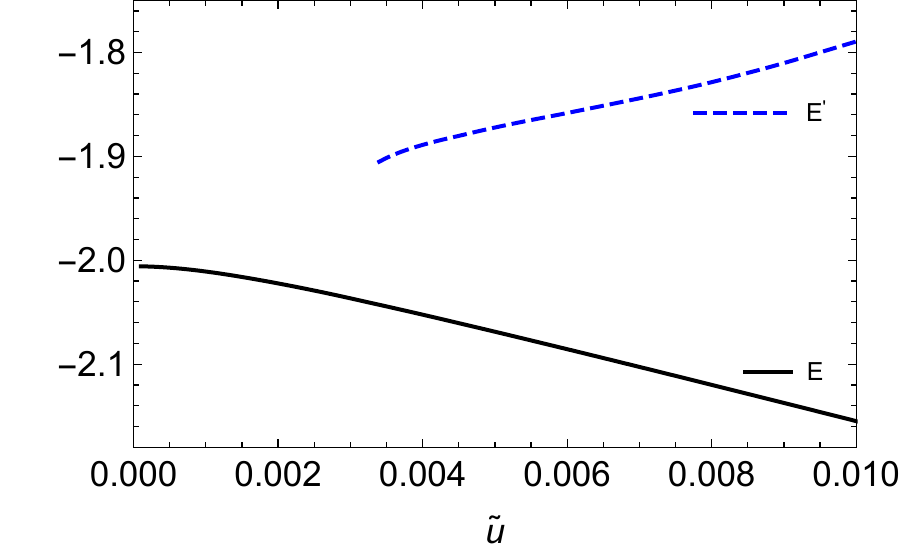}
\caption{
{ {\bf Scaling of the stable and metastable states with electron--phonon coupling strength, $\tilde{u}$.} $E$ is the ground-state energy, $E'$, is the energy of the additional self-consistent solution of the equation $E'=E_\mathsf{JM_J}-\Sigma_{\mathsf{JM_J}}(E')$, which is found at energies above $\text{Max}\left[-\text{Re}\,\Sigma_\mathsf{JM_J}(E')\right]$ (see Fig.\ref{graphicalsolution}). An approximately linear scaling with $\tilde{u}$ is found for both states, yielding a linear dependence of $\omega=E'-E$ on $\tilde{u}$ in the parameter range investigated.}
\label{scaling}}
\end{figure}
\end{small}

\end{document}